\documentstyle[12pt]{article}
\begin{document}
\input psfig

\title{Soliton propagation on a gravitational plane-wave collision
spacetime}

\author{ Gabriel B. Nagy, Reinaldo J. Gleiser\thanks{Researcher of
CONICET} and  Andres D. Dagotto\thanks{Current address: IBM of
Argentina}}

\date{}

\maketitle

\begin{center}
Facultad de Matem\'atica Astronom\'{\i}a y F\'{\i}sica \\
Universidad Nacional de C\'ordoba \\
Dr. Medina Allende y Haya de la Torre, \\
Ciudad Universitaria, \\
(5000) C\'ordoba, Argentina.
\end{center}

\begin{abstract}
We present a new family of exact solutions of the Einstein equations
that may be interpreted as representing the propagation of a pair of
solitons, in the background of a plane-wave collision spacetime. The
family is constructed through the Khan-Penrose procedure, as an
extension  of a known metric from an interaction region. The metric in
the interaction region is obtained as a diagonal solitonic perturbation
of Rindler's spacetime, applying the Belinskii and Zakharov Inverse
Scattering Method (ISM), with two real poles and one complex pole and
its complex conjugate. We use in this solution a non-standard
renormalization procedure, obtaining  solutions that contain two more
parameters than the analogous solution that results applying the
standard ISM.

We analyze the asymptotic behaviour of this family of solutions in the
limit where the determinant of the two by two Killing part of the
metric vanishes. We find that there exists a curvature singularity in
this limit, except when the free parameters contained in the solutions
satisfy certain relation, in which the new parameters introduced by the
non-standard renormalization procedure play an essential role. When
this condition is satisfied, it is possible to find a transformation to
a coordinate system where the metric is regular in the limit indicated
above, and we show that the resulting collision spacetimes contain in
that region a Killing Cauchy horizon instead of a curvature
singularity, as in the general case. Finally, we analytically extend
this subfamily through the horizon, and we find a curvature singularity
in this extension, that may be considered as the result of the
perturbation introduced in the interaction region by to the presence
and propagation of the two complex poles.

\end{abstract}

\section{Introduction}

The global structure of a gravitational plane-wave collision spacetime,
and, in particular, the singularities produced after the collision
because of the strong mutual focusing, is a subject of permanent
interest in the area of exact solutions in general relativity.  The
first exact vacuum solution of this type was obtained by Khan and
Penrose\cite{kp}. It describes the head-on collision of two parallel
polarized\footnote{This is equivalent to saying that the $2\times 2$
part of the metric corresponding to the plane symmetry is diagonal.}
impulsive plane waves. The final result of the collision is a curvature
singularity that can not be avoided by any observer (in this sense we
can say that the singularity is spacelike). This type of singularity
was also obtained by Nutku and Halil in \cite{nh}, where they describe
the collision of two impulsive plane waves with arbitrary relative
polarization. 

Taking a different starting point, Szekeres \cite{s1,s2} analyzed
the collision of plane waves as an initial value formulation, obtaining
an exact solution corresponding to the collision of two parallel
polarized but otherwise arbitrary plane waves. A large set of solutions
was obtained by Chandrasekhar, Ferrari and Xanthopoulos, through the
use of relations between the mathematical theory of black holes and
that of plane-wave collisions \cite{cf,cx1,cx2,cx3,cx4}. The crucial
point to obtain these solutions is that, in both cases, Einstein's
equations are reducible to the same Ernst equation. In all these
solutions, a curvature singularity develops as the result of the
gravitational interaction, except in \cite{cx3} where a Killing Cauchy
horizon is obtained, instead of the curvature singularity. This horizon
is a 3-dimensional hypersurface where
the spacetime plane-symmetry is broken, due to the fact that one of the
two spacelike Killing vectors becomes null on the horizon.
Chandrasekhar finds an analytic extension through this horizon,
verifying that the null Killing vector becomes timelike through the
horizon. He also finds a timelike curvature singularity in the extended
region. This kind of behaviour, that is, the development of a curvature
singularity, or the break-down of the plane symmetry by the formation
of a Killing Cauchy horizon, is the general outcome of an arbitrary
plane-wave collision. This result was proved by Tipler \cite{t}, and
his arguments suggest that these singularities are peculiar features of
plane waves, because singularities are also the consequence of a
collision of self-gravitating plane waves of other fields with
arbitrary small energy density. The stability of the Killing Cauchy
horizon upon variations keeping the plane symmetry \cite{y1}, and the
asymptotic behaviour of a parallel-polarized plane-wave collision
solutions near the singularity \cite{y2}, were analyzed by Yurtsever. In
this last work Yurtsever concludes that the metric near the singularity
is asymptotic to a Kasner solution.
 
As already mentioned, the solution found by Szekeres was obtained
through an initial value formulation. This procedure, however, presents
the difficulty of having to solve this initial value formulation in
general relativity which, even in the highly symmetric cases considered
in plane-wave collision analysis, is a non-trivial task. This is
probably the reason why most of the research that has provided exact
solutions to the problem has been based on the so-called Khan-Penrose
construction (KPC) \cite{kp}. In this construction the initial value
formulation is side-stepped, and the procedure acts as a generating
technique which, starting from a {\it known solution} to Einstein's
equations, not necessarily related to the collision problem but
satisfying certain conditions \cite{dgn1}, provides a full solution to
the problem.  This simplicity in the construction has the inconvenient,
however, that one does not know in advance what are the initial
conditions for which one has found a solution. Rather, the final
outcome of the collision is given, and one has to interpret the initial
conditions through an analysis of the solution.

Essentially, in the KPC the spacetime is divided into four regions
(labeled I, II, III, IV), separated by null hypersurfaces containing
the shock fronts, in such a way that IV is flat, II and III represent
the plane waves before the collision and I gives the interaction region
after the collision. A previously known solution provides the
interaction region I, and then it is suitable extended to the other
regions \cite{kp,dgn1}. As we shall see below, the metric may have, in
general, discontinuous derivatives on the hypersurfaces where the
extension is carried out, and this brings in the possibility of Dirac's
$\delta$-type singularities in the curvature tensor on these
hypersurfaces. Depending on whether some energy conditions are
satisfied, these singularities may be interpreted as corresponding to a
surface layer of null dust (massless particles). Nevertheless, they
should be absent if we want to restrict to vacuum solutions. The
corresponding conditions were found in \cite{dgn1}, and will be
described below, when we verify them in the new solution we present
here.

The starting point in the construction of our plane-wave collision
spacetime is an already known solution of Einstein's equations
\cite{dgp}, obtained with the Belinskii-Zakharov Inverse Scattering
Method (ISM) \cite{bz1}. This method provides a way of obtaining solutions
of Einstein's equations that can be regarded as solitonic perturbations
of a given seed metric. These solitonic perturbations display some of
the features found in solitons of other non-linear system such as the
Korteweg-de Vries (KdV) equation. For instance, at a given time the
perturbations is concentrated in certain region and travels with a well
defined speed. Specifically, the solitonic solutions are characterized
by the number of real or complex poles introduced as a part of the
generation method \cite{bz1}. Complex poles (which appear always
together with their complex conjugate to guarantee a real metric) are
associated with the type of solitonic perturbation mentioned above,
while real ones are associated with singularities in the resulting
metric (the nature of these singularities has been analyzed in
\cite{g}, \cite{dg}).  An important distinction regarding the solitonic
solutions we use here, is that they were constructed with a
non-standard renormalization procedure, given in \cite{dgp}, and,
therefore they contain two extra real free parameters per soliton, as
compared with the standard procedure. These extra free parameters play
a crucial role in the determination of the global properties of the
collision spacetimes.

In this paper, we take for region I a family of solutions obtained in
\cite{dgp}, which are solitonic perturbations of a seed metric given by
Rindler spacetime, containing two real and a pair of complex conjugate
poles.  This choice was taken because each real pole provides a
singularity that will be interpreted as a shock front in the plane-wave
collision spacetime, needed by the KPC (see below), and each complex
pole gives extra non-trivial gravitational structure that contributes
to the interaction after the collision. This extra structure and the
new free parameters due to the non-standard renormalization procedure
are the main new features of this family of solutions, as compared with
earlier work of Ferrari et.al. \cite{fib}.  For simplicity we first
describe the resulting plane-wave collision spacetime with only two
complex poles (and of course, the two real poles needed for the KPC),
and then we generalize this spacetime to an arbitrary number of complex
poles.  The choice in the seed metric is inspired in the already
indicated asymptotic results obtained by Yurtsever \cite{y2}, who
proved that in certain cases the plane-wave collision spacetimes
evolves near the singularity (in the interaction region I) to a Kasner
metric, of which the Rindler spacetime is a particular case. The
particular Rindler case is very interesting because of the fact that
the curvature singularity that characterizes a Kasner metric is changed
to a Killing Cauchy horizon for the Rindler spacetime. It is then
naturally interesting to see if the solitonic perturbations of
Rindler's spacetime that can be interpreted as the propagation of
solitons on a plane-wave collision spacetimes, maintain this horizon or
if, instead, in all cases they lead to the development of a curvature
singularity.

In Section 2 we present the solitonic family of metrics, solutions of
Einstein's equations, and we analyze the possibility of performing the
KPC. We find that this construction is possible only for certain values
of the free parameters of the family. We also confirm that outside this
range, the resulting spacetime cannot be interpreted as a collision
spacetime.  In Section 3 we compute the asymptotic expression for the
collision metric near the singularity in the interaction region I. With
this asymptotic metric we calculate the Kretschmann scalar and we find
a relation between the parameters that define the family, such that
this scalar does not diverge. This result suggests the possibility of
the presence of a Killing Cauchy horizon instead of the curvature
singularity. This suggestion is confirmed in Section 4, by finding a
coordinate transformation which, for the cases mentioned above, lets us
extend analytically the collision spacetime through the horizon. We
also prove that a new curvature singularity is present in this
analytically extended region, related to the presence and propagation
of complex poles in the original solution.

\section{The plane-wave collision spacetime}

The starting point in the construction of a collision spacetime through
the KPC, is a solution of Einstein's equations admitting two commuting
spacelike Killing vectors. It is well known that the metric for this
type of spacetimes can be written in the form
\begin{equation}
ds^2 = f(u,v) \; du\,dv + g_{ab}(u,v) \; dx^adx^b
\label{genmetric}
\end{equation}
where indices $a,b$  can take values $1,2$ while $x^a$ denotes the
Killing coordinates, and $u,v$ are null coordinates.  We define the
region I of the collision spacetime, or interaction region, as the set
of events with $u>0$, $v>0$ and the metric given by (\ref{genmetric}).
The set of events given by $u>0$, $v<0$ is region II, while $u<0$,
$v>0$ defines region III, and finally $u<0$, $v<0$ corresponds to
region IV. The coefficients of the metric in regions II, III and IV are
defined in terms of metric in region I as follows
$$
\begin{array}{rclcrcl}
f^{(II)}(u,v) & =& f^{(I)}(u,0) & \hspace{0.5cm} & g^{(II)}_{ab}(u,v)
&=& g^{(I)}_{ab}(u,0) \\
f^{(III)}(u,v) & =& f^{(I)}(0,v) & \hspace{0.5cm} & g^{(III)}_{ab}(u,v)
&=& g^{(I)}_{ab}(0,v) \\
f^{(IV)}(u,v) & =& f^{(I)}(0,0) & \hspace{0.5cm} & g^{(IV)}_{ab}(u,v)
&=& g^{(I)}_{ab}(0,0)
\end{array}
$$
where $f^{(K)}$ and $g^{(K)}_{ab}$ are the metric coefficients in
region $K=I, II, III, IV$. The metric thus obtained is well defined and
continuous in a neighbourhood of $u=0$, $v=0$. It can be checked that
if the metric in region I corresponds to a vacuum solution, then the
metric in regions II, III, and IV is also a vacuum solution. In fact,
in IV the metric is flat while in II and III the metric coefficients
depend on only one null coordinate. The KPC is therefore interpreted as
the head-on collision of two gravitational plane waves, with region I
representing the interaction region after the collision. It is clear
that the metric may have, in general, discontinuous derivatives on the
hypersurfaces $u=0$ and $v=0$, and this implies the possibility of
Dirac's $\delta$-type singularities on the curvature tensor,
which should be absent if we want vacuum solutions. It can
be checked that a sufficient condition to avoid this singular behaviour
of the curvature tensor is
\cite{dgn1}
\begin{equation}
\lim_{v \to 0^+} \alpha_{,v} =0 \hspace{0.5cm} \lim_{u \to 0^+}
\alpha_{,u} =0
\label{cond}
\end{equation}
where $\alpha \equiv \sqrt{|g_{ab}|}$ (i.e. the square root of the
determinant of the Killing part of the metric) and subindices in
(\ref{cond}) indicate differentiation.

We choose as the metric that represents the interaction region an
already known solution of Einstein's equations \cite{dgp}, obtained by
applying the ISM \cite{bz1} to a seed metric given by the Rindler
spacetime. We choose this spacetime as the background for the solitonic
perturbation for the following reasons. First, it is of interest to search
for collision spacetimes that do not develop curvature singularities,
and, since a collision spacetime evolves near the singularity in the
interaction region asymptotically to a Kasner spacetime \cite{y2}, we
are rather naturally lead to study the perturbations of a Rindler
metric, a special case of a Kasner spacetime, which does not develop a
curvature singularity.  Second, we consider a solitonic perturbation
that incorporates also complex poles, in order to analyze the effect of
this new structure and its propagation, over that of the already known
solutions \cite{fib}, where only real poles were included.

The chosen family of solution was obtained applying a non-standard
renormalization procedure \cite{dgp}, that adds two extra real
parameters, $\delta$ and $\gamma$, to those already present in the
standard Belinskii-Zakharov ISM.  The expression for the metric in
region I is,
$$
ds^2 = f(t,z)(dz^2 - dt^2) + G_{11}(t,z) \; dx^2 + G_{22}(t,z) \; dy^2
$$
where the Killing coordinates are now $x$ and $y$, and the functions
$f(t,z)$, $G_{11}(t,z)$ and $G_{22}(t,z)$ are given by
\begin{eqnarray}
G_{11}(t,z) &=& t^{2\delta} \left( \prod_{k=1}^{n} \mu_k
\right)^{\gamma}
\label{def1} \\
G_{22}(t,z) &=& \frac {t^{2}}{G_{11}(t,z)} 
\label{def2}\\
f(t,z) &=&  C \; t^{2\delta(\delta-1)}  \left( \prod_{k=1}^{n} \mu_k
\right) ^{(2\gamma^{2} + 2\gamma\delta -\gamma)}  \left(  \frac{
\prod_{k > j=1}^{n} (\mu_k-\mu_j)^2 }{ \prod_{k=1}^{n} ( (\mu_k)^2 -
t^2 ) } \right) ^{\gamma^2} 
\label{def3} \\
\mu_k(t,z) &=& \left( (W_k-z)+\epsilon_k \sqrt{(W_k-z)^2-t^2} \right)
           \hspace{1cm} (k=1\cdots n)
\label{def4}
\end{eqnarray}
where $\epsilon_k \pm 1$, $n=4$, $W_1=z_1$, $W_2=z_2 +i \omega$,
$W_3=z_2 - i\omega $ and $W_4=z_3$, with
$\omega$, $z_1$, $z_2$, $z_3 $,  and $C$ real constants. We further
assume that $z_1 < z_2 < z_3$ to ensure that
the perturbation produced by the complex poles can be observed inside
the interaction region.  Summarizing, assuming that we
keep the values of $\omega$, $z_1$, $z_2$, $z_3 $,  and $C$ fixed, for
each choice of the $\epsilon_k$ we have a well defined two-parameter
family of solutions, (given by  $\delta$ and $\gamma$), for the
following ranges of the coordinates (which represent a triangle in
$(z,t)$ coordinates)
\begin{equation}
\begin{array}{rcccccl}
-\infty & < & x,y & < & \infty & & \\
z_1 & < & z & < & z_3 & &  \\
z_1-z & < & t & , & z-z_3 & < & t\hspace{0.5cm}
\mbox{and}\hspace{0.5cm} t<0.
\end{array}
\label{chart}
\end{equation}
With the choice of coordinates in (\ref{def1})-(\ref{def3}) we have
that $\alpha=t$ (this is always possible  in metrics of the form
(\ref{genmetric})). The constants $W_k$ are directly related with the
poles in the ISM \cite{bz1}. The complex conjugate pair of poles
related to  $W_2$ and $W_3$ introduce a non-trivial gravitational
structure on the incoming plane waves, but do not give rise to a
divergent behaviour on the metric components. On the other hand, the
real poles related to $W_1$ and $W_4$, besides their contribution to the
non-trivial gravitational behaviour of the incoming plane waves,
introduce also a divergent behaviour in the function $f$.  This
function  is singular for $t=z_1-z$ and $t=z-z_3$ because of the
vanishing factors $(\mu_1^2 -t^2)^{\gamma^2}$ and $(\mu_4^2
-t^2)^{\gamma^2}$ in its denominator.

The KPC can be performed if there exist two null hypersurfaces (e.g.
$u=0$ and $v=0$) such that (\ref{cond}) holds, or, written in our
coordinate system, $\lim_{u\to 0^+} t_{,u} =0$ and  $\lim_{v\to 0^+}
t_{,v} =0$. Since any coordinate transformation from $(z,t)$ to a pair
of null coordinates $(u,v)$ must be of the form $t=F_1(u)+F_2(v)$,
$z=F_1(u)-F_2(v)$, then (\ref{cond}) implies that $\lim_{u\to 0^+}
F_{1,u} =0$ and  $\lim_{v\to 0^+} F_{2,v} =0$. But the jacobian of this
transformation is $2F_{1,u}F_{2,v}$, and, therefore, (\ref{cond})
implies that the coordinate transformation must be singular for $u=0$
and $v=0$. Therefore, since the metric in the null coordinates must be
well behaved on these hypersurfaces (a requirement of the KPC), then it
must be singular in the chart $(z,t)$,  on the same hypersurfaces.

Thus, a necessary condition to construct a collision spacetime from
(\ref{def1}) - (\ref{def3}) is to choose two $W_k$, say $W_1$ and
$W_4$, real, so that the metric is singular on the null hypersurfaces
$t=z_1-z$ and $t=z-z_3$, where the wave fronts will be located. The
existence of these singularities is not a sufficient condition for the
construction of a collision spacetime, because we must also explicitly
find the singular coordinate transformation to null coordinates that
ensures (\ref{cond}). We can determine this coordinate transformation
from the behaviour of the function $f$ near any of these two null
hypersurfaces. For example, near $t=z_1-z$ we have
$$
f(t,z)=\frac{f_0(t,z)}{(t+z-z_1)^{\gamma^2/2}}
$$
with $f_0(t,z)$ finite and different from zero near the hypersurface.
Near the other hypersurface we have an analogous expression. It can now
be checked that the following is the desired coordinate transformation
\begin{equation}
\begin{array}{lcl}
u=(z-z_1+t)^{1/\sigma} &  \;\;\; , \;\;\; & u \in
(0,(z_3-z_1)^{1/\sigma}] \\
v=(z_3-z+t)^{1/\sigma} &  , & v \in
(0,(z_3-z_1)^{1/\sigma}]
\end{array} 
\label{transf}
\end{equation}
with
\begin{equation}
\sigma \equiv {1 \over 1-\gamma^2/2} \;\;\; ,\;\;\;
(u^{\sigma}+v^{\sigma})  -(z_3-z_1) < 0 \;\;\; , \;\;\;
\gamma^2 < 2
\end{equation}
because under this transformation the metric is changed to
$$
ds^2= -\sigma^2 {f(u,v) \over (u v)^{1-\sigma}} \; dudv + G_{11}(u,v)
\; dx^2 + \frac{t(u,v)^2}{G_{11}(u,v)} \; dy^2
$$
and the divergent behaviour coming from the factors $({\mu_1}^2
-t^2)^{-\gamma^2}$ and $({\mu_4}^2 -t^2)^{-\gamma^2}$ in (\ref{def1}) -
(\ref{def3}) is cancelled by the jacobian of the transformation, and
 the metric is well behaved on these hypersurfaces.  Finally, to
carry out the KPC, it is necessary to check that (\ref{cond}) holds. It
is easy to see that
$$
t=\frac{1}{2} \left[ (u^{\sigma} + v^{\sigma}) -(z_3-z_1) \right]
$$
so we have that
\begin{equation}
\lim_{u \to 0^+} t_{,u} = \lim_{u \to 0^+}\frac{\sigma}{2} \;
u^{\sigma-1} =0 \Longleftrightarrow \gamma^2 < 2
\label{gl2}
\end{equation}
and the same condition for $\gamma$ is found on $v=0$.  Summarizing, we
propose as the interaction region of our plane-wave collision
spacetime, a family of metrics obtained as solitonic perturbation of
Rindler's spacetime, which necessarily have two real poles to carry out
the KPC.  The construction is performed by obtaining explicitly new
null coordinates $u,v$ such that the wave fronts are the hypersurfaces
$u=0$ and $v=0$, and the metric has continuous first derivatives across
these wave fronts.  This construction, however, is not possible for the
whole family of solitonic metrics, but is restricted to the subfamily
with $\gamma^2 <2$. 

The rest of the solitonic family, for $\gamma^2 \geq 2$, cannot be
considered as a collision spacetime, because the regions $t=z_1-z$ and
$t=z-z_3$, which we would like to interpret as the wave fronts are, in
fact, at past null infinity, as can be seen from the following
argument. First, we perform the following coordinate
transformation\footnote{This coordinate transformation is only valid
for $\gamma^2 > 2$, but it is easy to check that the same ideas are
applicable in the case of $\gamma^2 = 2$, with a coordinate
transformation of the form $u=\ln (z-z_1+t)$, $v=\ln (z_3-z+t)$;
arriving at the same conclusion as in the case $\gamma^2 > 2$}.
\begin{equation}
\begin{array}{lcl}
u=(z-z_1+t)^{-1/\sigma^{\prime}} &  \;\;\; , \;\;\; & u \in
((z_3-z_1)^{-1/\sigma^{\prime}},\infty) \\
v=(z_3-z+t)^{-1/\sigma^{\prime}} &  , & v \in
((z_3-z_1)^{-1/\sigma^{\prime}},\infty)
\end{array} 
\label{transf1}
\end{equation}
with
\begin{equation}
\sigma^{\prime} \equiv {1 \over \gamma^2/2 -1 } \;\;\; , \;\;\;
(\frac{1}{u^{\sigma^{\prime}}}+\frac{1}{v^{\sigma^{\prime}}})
-(z_3-z_1) < 0 \;\;\; , \;\;\; \gamma^2 > 2
\end{equation}

Notice that with $u$, and $v$, defined as in (\ref{transf1}), the
hypersurfaces $t=z_1 -z$ and $t=z-z_3$ correspond, respectively, to
$u=\infty$ and $v=\infty$. The null vectors $(\partial / \partial u)^a$
and $(\partial / \partial v)^a$ are, therefore, {\em past} directed.

Next, we consider a future directed  null geodesic $\Gamma(u)$, with
tangent vector $T^a \equiv - (\partial / \partial u)^a$, that contains
an arbitrary point in region (\ref{chart}), whose new coordinates are
$(u_0,v_0)$. The coordinate $u$ is not an affine parameter for this
geodesic, because
$$
T^a\nabla_aT^b = -\Gamma^u_{uu} T^b
$$
but, solving  
$$
\frac{d\lambda}{du} = \exp \left(-\int_{u_0}^u \Gamma^u_{uu}(\tau) \;
d\tau \right).
$$
we may construct an affine parameter $\lambda$ such that $T^a \equiv
(\partial / \partial \lambda)^a$.  For a metric of the form
(\ref{genmetric}), in coordinates given by (\ref{transf1}), we have

$$
\Gamma^u_{uu} = \ln \left( \frac{f}{u^{(\sigma^{\prime} +
1)}}\right)_{,u}
$$
and this implies that
$$
\lambda(u) = -\int_{u_0}^u \frac{f(U,v_0)}{U^{(\sigma^{\prime}+1)}} dU.
$$
We know that the function $f(u,v)$ appearing in the metric has the form
$f(u,v)=f_0(u,v)\;u^{(\sigma^{\prime}+1)}$ near the hypersurface
$t=z_1-z$, with $f_0$ finite and different from zero in the
hypersurface. This asymptotic behaviour of $f$, together with the fact
that when $\gamma^2 > 2$ the hypersurface $t=z_1-z$ corresponds to
$u=\infty$, means that the affine parameter of the geodesic on this
hypersurface is
$$
\lambda(\infty) = -\int_{u_0}^\infty f_0(U,v_0) \; dU = -\infty.
$$
Thus, the geodesic reaches the hypersurface with infinite affine
parameter implying that this hypersurface is at past null infinity and,
therefore, the solitonic spacetime cannot be considered as a part of 
a collision spacetime. On the other hand, for  $\gamma^2 < 2$, the
hypersurface that defines the boundary between regions I and III,
corresponds to $u=0$, and a similar argument shows that in this case a
past directed null geodesic reaches this boundary with a finite affine
parameter.
\begin{figure}
\psfig{figure=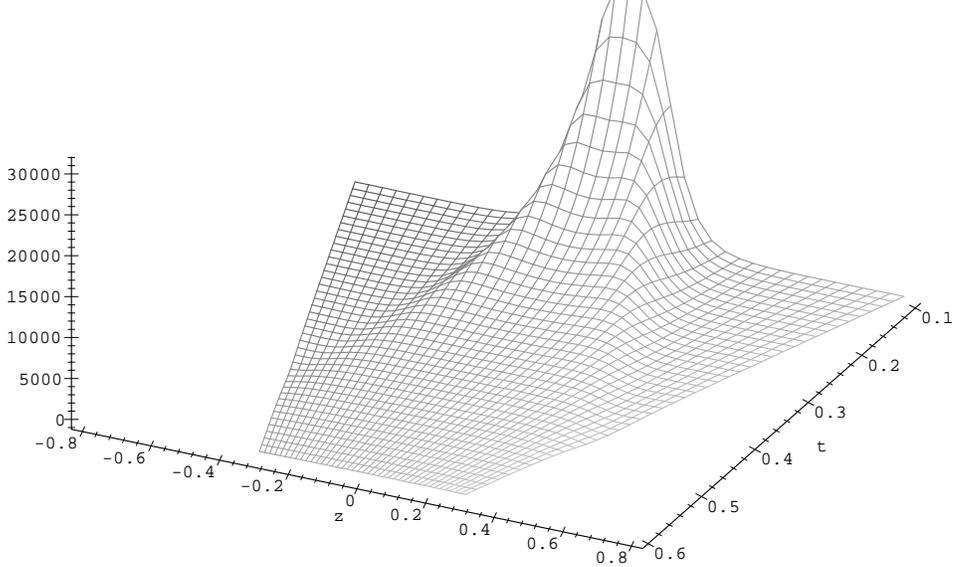}
\caption{The ratio $(\Psi_2)_{(4)} / (\Psi_2)_{(2)}$ in the
interaction region, for the parameters $\epsilon_1= \epsilon_2= 1$,
$\epsilon_4=-1$, $\omega =0.1$, $\delta=0$ and $\gamma=1$.}
\end{figure}

Finally, we are interested in comparing the geometric structure that
results from the presence of the complex soliton poles, with that
obtained in the case where these complex solitons are absent
\cite{fib}. This comparison can be done through the study of the Weyl
scalars.  The complexity of the metric coefficients makes this analysis
very difficult, so we concentrate our attention on the Weyl scalar
$\Psi_2$, which is different from zero only in the interaction region.
This Weyl scalar has the properties that it involves only first
derivatives of the metric functions, and that it is the only non-zero
Weyl scalar which is invariant under simultaneous rescaling of the
tetrad vectors $l^a$ and $n^a$. So, it is  plausible to think that if
$\Psi_2$ is singular, the metric will be singular.
\begin{figure}
\psfig{figure=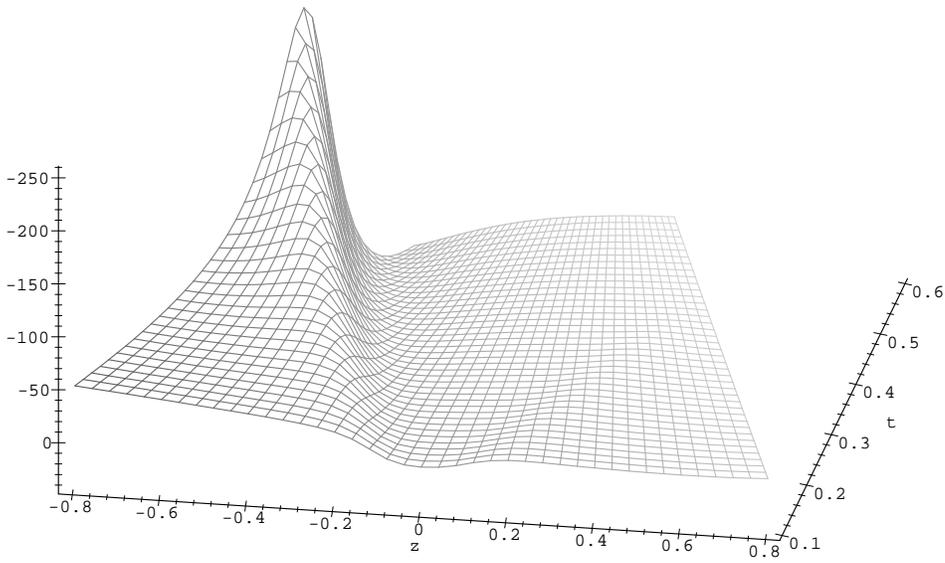}
\caption{The ratio $(\Psi_2)_{(4)} / (\Psi_2)_{(2)}$ in the
interaction region, for the parameters $\epsilon_1= \epsilon_2 =
\epsilon_4 = 1$, $\omega =0.1$, $\delta=0$ and $\gamma=1$.}
\end{figure}

The comparison is performed by introducing the following null tetrad,
defined in the interaction region,
\begin{eqnarray*}
l^a &=& \frac{1}{\sqrt{2 f(t,z)}} \left[ \left(
\frac{\partial}{\partial t}\right)^a + \left( \frac{\partial}{\partial
z} \right)^a \right] \\
n^a &=& \frac{1}{\sqrt{2 f(t,z)}} \left[ \left(
\frac{\partial}{\partial t}\right)^a - \left( \frac{\partial}{\partial
z} \right)^a \right] \\
m^a &=& \frac{1}{\sqrt{2}} \left[ \frac{1}{\sqrt{g(t,z)}} \left(
\frac{\partial}{\partial x}\right)^a +i \; \frac{\sqrt{g(t,z)}}{t}
\left( \frac{\partial}{\partial y} \right)^a \right] \\
\overline{m}^a &=& \frac{1}{\sqrt{2}} \left[ \frac{1}{\sqrt{g(t,z)}}
\left( \frac{\partial}{\partial x}\right)^a - i \;
\frac{\sqrt{g(t,z)}}{t} \left( \frac{\partial}{\partial y} \right)^a
\right] 
\end{eqnarray*}
and computing the corresponding expressions for the Weyl scalar
$(\Psi_2)_{(4)}$ of the four-soliton metric and $(\Psi_2)_{(2)}$ of the
corresponding two real soliton metric, that is, a metric with the same
parameters as the former, but without the inclusion of the two complex
solitons. In Figures 1 and 2, we plot the ratio $(\Psi_2)_{(4)} /
(\Psi_2)_{(2)}$, for some particular choices of the free parameters. We
can see that the relative behaviour of these two of metrics depends in
a crucial way on the choice of these free parameters. In Figure 1 we
can see that $(\Psi_2)_{(4)}$ becomes very different from
$(\Psi_2)_{(2)}$ in the region $t \to 0$, while in Figure 2 we see that
this difference disappears in the same limit. The propagating nature of
the soliton perturbations is also clear in both figures.

\section{Asymptotic behaviour}

Once we have constructed the plane-wave collision spacetime from the
solitonic subfamily of Einstein's equations solutions, the next natural
step is to analyze the resulting spacetime, in particular, in the
interaction region. It is well known \cite{t} that as the result of the
strong mutual focusing of the waves, we have in this region the
development of either a curvature singularity, or a breaking of  the
plane symmetry, with the creation of a Killing Cauchy horizon. In the
solution we  present here, this singular behaviour appears in the limit
$t \to 0^-$ in the chart $(t,z)$. The exact explicit expressions for
the coefficients of the metric in that region are very complicated. On
this account, and to simplify the analysis, we consider appropriate
Taylor series expansions, instead of the exact expressions. We use
these expansions to compute the Kretschmann scalar, looking for
sufficient conditions to ensure a curvature singularity. As we shall
see, for some choices of the free parameters $\delta$, $\gamma$,
$\epsilon_k$, $(k=1\cdots 4)$, this scalar diverges in the limit $t \to
0^-$, indicating a curvature singularity, but we find a finite limit
for other choices of these parameters. We can show that for the latter
spacetimes the metric approximates a Rindler form, suggesting that an
appropriate coordinate transformation  may be used to prove that the
singular limit $t \to 0^-$ in the chart $(t,z)$ is only a coordinate
singularity. In fact, using this coordinate transformation we find that
the hypersurface corresponding to $t=0$ is indeed a Killing Cauchy
horizon, and that the metric may be extended through this horizon.
This extension will be carried out in the next Section, while in this
Section we construct the required coordinate transformation.

We first calculate the expansion for the metric coefficients in the
limit $t \to 0^-$ and $z = \mbox{const}$. This may be done for
arbitrary number of solitons $n$, without the restriction to $n=4$. We
begin our study expanding the coefficients $\mu_k$ near $t \sim 0$
$$
\mu_k(t,z) \simeq \left[ (W_k-z) + \epsilon_k \sqrt{(W_k
-z)^2}\right] - \frac{\epsilon_k t^2}{2\sqrt{(W_k-z)^2}}
$$
with $k=1\cdots n$. Here we have to consider different possibilities.
If $W_k$ is complex, we have,
$$
\mu_k(t,z) \simeq \left\{
\begin{array}{lcrcl}
 2\sqrt{(W_k-z)^2} & \mbox{   if   } &\epsilon_k &=& 1 \\
 (t^2/2)/\sqrt{(W_k-z)^2} & \mbox{   if   }  & \epsilon_k &=&
 -1
\end{array} \right.
$$
while for $W_k$ real, we have only two possibilities, say, $W_1=z_1$ and
$W_4=z_3$. If we recall that $z_1 < z < z_3$ we obtain
\begin{eqnarray*}
\mu_1(t,z) &\simeq &\left\{
\begin{array}{lcrcl}
  (t^2/2)/(z_1-z) & \mbox{   if   }  & \epsilon_1 
  &=& 1 \\
 2(z_1-z) & \mbox{   if   } &\epsilon_1 &=& -1 \\
 \end{array} \right. \\
\mu_4(t,z) &\simeq &\left\{
\begin{array}{lcrcl}
 2 (z_3-z) & \mbox{   if   } &\epsilon_4 &=& 1 \\
 (t^2/2)/(z_3-z) & \mbox{   if   }  & \epsilon_4 &=&
 -1
\end{array} \right.
\end{eqnarray*}

Defining $B_{\tilde{k}}(z) \equiv +2 \sqrt{(W_k-z)^2}$, $(\tilde{k}
\neq 1,\;  4)$, $B_1(z) \equiv2 (z_1-z)$ and $B_4(z) \equiv 2 (z_3-z)$,
we may summarize all these results as
$$
\mu_k(t,z) \simeq \left\{
\begin{array}{c}
 B_k(z)  \\
  t^2/B_k(z)
\end{array} \right. .
$$

Let us assume that we have chosen the $\epsilon_k$ such that there are
$m$ functions $\mu_k(t,z)$ which behave as $t^2/ B_k(z)$ when $t \to
0^-$, $z=\mbox{const}$, and $n-m$ such that $\mu_k(t,z) \to B_k(z)$ in
the same limit.  Then, the first order of the asymptotic expression for
the metric coefficients in this limit is the following
\begin{eqnarray*}
g(t,z) &\simeq & t^{2(\delta+m\gamma)} \left[ \frac{\prod_{j=m+1}^n
B_{k_j}(z)}{\prod_{i=1}^m B_{k_i}(z)}\right]^{\gamma} \\
f(t,z) &\simeq & C_1 \; t^{2(\delta+m\gamma)(\delta+m\gamma -1)} \left[
\frac{ \prod_{j=m+1}^n B_{k_j}(z)}{\prod_{i=1}^m B_{k_i}(z)}
\right]^{(2(\delta+m\gamma)-1)\gamma}
\end{eqnarray*}
where $C_1$ is a constant. Defining the function 
$$
\widetilde{g}(z) \equiv \left[ \frac{ \prod_{j=m+1}^n
B_{k_j}(z)}{\prod_{i=1}^m B_{k_i}(z)} \right]^{\gamma}
$$
and the constant $\phi \equiv (\delta+m\gamma)$, we may rewrite the
asymptotic form of the metric as follows
\begin{equation}
ds^2 \simeq C_1 \; t^{2\phi(\phi -1)} \;
\widetilde{g}(z)^{(2\phi-1)} \; (-dt^2 + dz^2) +
t^{2\phi} \; \widetilde{g}(z) \; dx^2 + t^{-2(\phi-1)}
\; \widetilde{g}(z)^{-1} \; dy^2.
\label{asymptmetric}
\end{equation}
With this expression for the metric, we obtain the following asymptotic
expression for the Kretschmann scalar,
$$
R_{abcd}R^{abcd} \simeq \frac{32 \phi^2 (\phi -1
)^2(\phi(\phi-1)+1) }{ C_1^2 \; t^{4\phi(\phi-1)}
\widetilde{g}(z)^{2(2\phi-1)} t^4}.
$$
We conclude that if the parameters $\delta$ and $\gamma$ are such that
$\phi$ is different from zero or one, then the corresponding plane-wave
collision spacetimes develop a curvature singularity for $t \to 0^-$.
On the other hand, if $\phi$ equals zero or one, we may prove that we
have  a coordinate singularity in the chart $(t,z)$ when  $t \to 0^-$,
by finding a coordinate transformation such that the metric is well
behaved in the same limit. To obtain the appropriate coordinate
transformation, we notice, from (\ref{asymptmetric}), that the
asymptotic form of the metric corresponds to a Rindler spacetime. For
example, for $\phi =0$ we have,
$$
ds^2 \simeq  \widetilde{g}(z)^{-1} \; (-dt^2 +  t^{2} \;   dy^2) +
\widetilde{g}(z)^{-1} \; dz^2 + \widetilde{g}(z) \; dx^2 
$$
where the constant $C_1$ was removed by a rescaling of the coordinates.
The singularity for $t=0$ corresponds here to the vanishing of the
factor $t^2$ in the coefficient of $dy^2$. But, if we introduce new
coordinates $\tau$ and $\eta$, defined by
\begin{equation}
\tau \equiv t \; \cosh (y) \hspace{1cm} \eta \equiv -t \; \sinh (y)
\label{transf1b}
\end{equation}
we have,
\begin{equation}
\-dt^2+t^2\; dy^2 = - d \tau^2 + d \eta^2  
\label{transf1a}
\end{equation}
and in this new coordinates the asymptotic
expression for the metric takes the following form
$$
ds^2 \simeq  \widetilde{g}(z)^{-1} \; (-d\tau^2 + d\eta^2) +
\widetilde{g}(z)^{-1} \; dz^2 + \widetilde{g}(z) \; dx^2
$$
and the singularity for $t=0$, which in this chart corresponds to $\tau
= \eta$,  is removed. A similar transformation may be applied if
$\phi=1$, with the change $y \leftrightarrow x$, to prove regularity
for $t=0$ when $\phi=1$. We can also check that one of the spatial
Killing vectors becomes null in this hypersurface. In the chart
$(t,z,x,y)$ the Killing vectors are
$$
\xi^a = \left( \frac{\partial}{\partial x}\right)^a \hspace{1cm}
\zeta^a = \left( \frac{\partial}{\partial y}\right)^a
$$
while in the chart $(\tau,\eta,z,y)$ they have the following expression
$$
\xi^a = -\eta \left( \frac{\partial}{\partial \tau}\right)^a  - \tau
\left( \frac{\partial}{\partial \eta}\right)^a \hspace{1cm}
\zeta^a = \left( \frac{\partial}{\partial y}\right)^a
$$
and then $\xi^a$ becomes null on the hypersurface $\tau = \pm \eta$. In
fact, it can be checked that this hypersurface is a Killing Cauchy
horizon. This is because this hypersurface has no border (its
generators are the integral lines of the null and spatial Killing
vectors $\xi^a$ and $\zeta^a$ respectively, and the spatial vector
$(\partial/\partial z)^a$ with $z$ varying up to the ``fold
singularity''); and this hypersurface is $C^{-1}$. The only non-trivial
step is to check the completeness of the null geodesic generator. But
it can be checked that the vector  $\xi^a$ is geodesic on the horizon,
and since it is a Killing vector, then is complete.

\section{Analytic Extension}

We proved in the previous Section that, for certain values of the
parameters $\delta$ and $\gamma$, the corresponding plane-wave
collision spacetimes develop a Killing Cauchy horizon after the
collision, but the metric is otherwise regular. This behaviour was
proved by writing the metric in the interaction region in appropriate
Rindler coordinates $(\tau,\eta,z,x)$, such that the interaction region
corresponds to  $ |\tau | > |\eta |$ with $\tau < 0$, and the horizon
is on the hypersurface $\tau = -|\eta |$. It is then possible to extend
the collision spacetime through the horizon, by analytically extending
the definition of the metric coefficients, that is, defining this
coefficients for all possible values of $\tau$ and $\eta$, where they
are regular.

The global structure of the spacetime  extended in this form
 may be separated in four regions. The first region, which we call
 ${\cal{C}}_1$, corresponds to the original interaction region of the
collision spacetime, that is $ |\tau | > |\eta |$ and $\tau < 0$; the
second, ${\cal{C}}_2$, and third, ${\cal{C}}_3$, correspond to $
|\tau | < |\eta |$ and $\eta < 0$ or $\eta >0$ respectively; and the
fourth, ${\cal{C}}_4$, is delimited by $ |\tau | > |\eta |$ and $\tau
> 0$.

However, the explicit expressions for the metric coefficients in this
Rindler coordinates are quite involved, and so, to study the existence
of singularities in any of those regions, it is convenient to perform
appropriate coordinates transformations in each region. For example in
${\cal{C}}_4$, it is convenient to make the transformation $t^2=
\tau^2 -\eta^2$, $\tanh (x) = -\eta /\tau$, which is the inverse of
(\ref{transf1b}). We obtain here the same functional expression for the
metric coefficients as in (\ref{def1})-(\ref{def3}), but now $t$ is
positive. We  therefore see that the analytic extension of the
plane-wave collision spacetime through the Killing Cauchy horizon
involves more than merely matching the $(t,z,x,y)$ charts for $t<0$ and
$t>0$, because we also have to consider the regions ${\cal{C}}_2$ and
${\cal{C}}_3$.

The analysis of the extended spacetime in the regions ${\cal{C}}_2$ and
${\cal{C}}_3$, is simplified by the following coordinate
transformation: $\tau = \tilde{t} \sinh (y)$, $\eta=\tilde{t} \cosh
(y)$. It can be checked that the metric coefficients obtained are the
same as those in (\ref{def1})-(\ref{def3}), but with the change $t^2 =-
\tilde{t}^2$.  This observation is useful in the analysis of the
behaviour of metric coefficients in these regions. It can be checked
that the coefficient $f(\tilde{t},z)$ is singular in the limit $z =z_2$
and $\tilde{t} \to \omega$, due to the vanishing of the factors
$(\mu_k(\tilde{t},z))^2 + (\tilde{t})^2) $ with $(k =2,3)$. Therefore,
this singular behaviour is directly related to the presence of a
complex pole in the ISM. To decide whether this singularity corresponds
to a curvature or a coordinate singularity, we again compute the
Kretschmann scalar in the limit $z =z_2$ and $\tilde{t} \to \omega$.
For technical reasons, it is easier to perform this calculation in
coordinates $\xi$, $\zeta$ given by $\tilde{t}^2 = \omega^2 (\xi^2 +1
)(1- \zeta^2)$ and $z = \omega \xi \zeta$, where for simplicity we
choose $z_2=0$. We calculate the asymptotic expression for the
Kretschmann scalar in the limit $\xi =0$ and $\zeta \to 0^-$, which it
is equivalent to the limit $z=0$ and $\tilde{t} \to \omega^-$, and we
obtain the following expression
$$
R_{abcd}R^{abcd} = \left( \frac{2 \gamma}{\sqrt{|F_0|} (1-\zeta^2)
}\right)^4 \zeta^{4(\gamma^2 -2)} \left[ 1 + O(\zeta ) \right]
$$
with $ F_0 = f_0(\xi=0,\zeta =0)$ and $f_0(\xi,\zeta)$ is defined as
$f(\xi,\zeta) \equiv f_0 (\xi,\zeta) / (\xi^2 + \zeta^2)^{\gamma^2}$,
and it can be checked that $f_0(\xi,\zeta)$ is the non singular part
of $f(\xi, \zeta)$ in the limit $\xi \to 0$, $\zeta \to 0$.  We
conclude that all these collision spacetimes, as defined in Section
2, have a curvature singularity in the limit $z =z_2$ and $\tilde{t}
\to \omega^-$, because, as we proved in Section 2, all these
collision spacetimes satisfy $\gamma^2 < 2$.

\goodbreak

\section{Conclusions}

We have presented a new family of plane-wave collision spacetime,
constructed from an interaction region, through the Khan-Penrose
procedure. The interaction region corresponds to a diagonal
solitonic perturbation of Rindler's spacetime, obtained applying the
ISM, with two real poles and a pair of complex conjugate poles. We
further use a renormalization procedure developed in \cite{dgp} to
obtain a family of solutions with two more parameters than the
analogous solution obtained with the standard ISM. We find conditions
on the free parameter that characterize the family of solutions such
that it is possible to perform the Khan-Penrose construction. We
conclude that only a restricted subfamily (given by $\gamma^2 < 2$),
can be considered as a plane-wave collision spacetime, because only in
this case the hypersurfaces corresponding to the gravitational wave
fronts can be reached by a congruence of past null geodesic, starting
somewhere in the presumed interaction region, with finite affine
parameter. For the rest of the family, these hypersurfaces are located
at past null infinity, precluding their interpretation as plane-wave
collision spacetimes.

We also study the perturbation introduced by the two complex poles to
the already known plane-wave collision solution corresponding only to
two real poles \cite{fib}. This perturbation is shown in Figures  1 and
2,  where we plot the Weyl scalar $\Psi_2$ of the four-soliton metric
divided by the same Weyl scalar of the two-soliton metric, for
different choices of the free parameters.

Next, we analyze the asymptotic behaviour of this family of solutions
in the limit where the determinant of the two by two Killing part of
the metric vanishes.  This analysis was carried out in the general case
of a metric with an arbitrary number $n>0$ of solitons. We find that
the Kretschmann scalar diverges, and therefore we have a curvature
singularity in this limit, except when the free parameters of the
family of solutions satisfy some relation, where the new parameters
introduced by the non-standard renormalization procedure are essential.
This condition also depends on the behaviour of the pole trajectory
functions $\mu_k(t,z)$, $(k=1 \cdots n)$, in this limit. When this
condition is satisfied, we find that the Kretschmann scalar is well
behaved, and studying the asymptotic expression of the metric
coefficients, that resembles that of Rindler's spacetime, we showed
that it is possible to find a coordinate transformation where the
metric is regular in the limit mentioned above. We proved that this
subfamily of collision spacetimes develops a Killing Cauchy horizon
instead of a curvature singularity.

Finally, we analytically extend this subfamily through the horizon. The
extended spacetime obtained contains two stationary regions with a
curvature singularity. This curvature singularity, which is timelike in
the sense that can be avoided by an observer, may be interpreted as 
resulting from the perturbation introduced in the interaction region by
the presence of the two complex poles.

\end{document}